# Experimental study of ionization yield of liquid xenon for electron recoils in the energy range 2.8 ÷ 80 keV

**D.Yu. Akimov** [a,b,*], **V.V. Afanasyev** [b], **I.S. Alexandrov** [a,b], **V.A. Belov** [a,b], **A.I. Bolozdynya** [b], **A.A. Burenkov** [a,b], **Yu.V. Efremenko** [c,b], **D.A. Egorov** [b], **A.V. Etenko** [d,b], **M.A. Gulin** [b], **S.V. Ivakhin** [b], **V.A. Kaplin** [b], **A.K. Karelin** [a,b], **A.V. Khromov** [b], **M.A. Kirsanov** [b], **S.G. Klimanov** [b], **A.S. Kobyakin** [a,e], **A.M. Konovalov** [a,b], **A.G. Kovalenko** [a,b], **A.V. Kuchenkov** [a,b], **A.V. Kumpan** [b], **Yu.A. Melikyan** [b], **R.I. Nikolaev** [b], **D.G. Rudik** [a,b], **V.V. Sosnovtsev** [b], **V.N. Stekhanov** [a,b]

[a] *SSC RF Institute for Theoretical and Experimental Physics of National Research Center "Kurchatov Institute", Moscow, 117218, Russian Federation*

[b] *National Research Nuclear University "MEPhI" (Moscow Engineering Physics Institute), 115409 Moscow, Russian Federation*

[c] *University of Tennessee, Knoxville, Tennessee 37996, USA*

[d] *National Research Centre "Kurchatov Institute", Moscow 123182, Russian Federation*

[e] *Moscow Institute of Physics and Technology (State University), Moscow, 117303, Russian Federation*
   *E-mail:* `akimov_d@itep.ru`

ABSTRACT: We present the results of the first experimental study of ionization yield of electron recoils with energies below 100 keV produced in liquid xenon by the isotopes: $^{37}$Ar, $^{83m}$Kr, $^{241}$Am, $^{129}$Xe, $^{131}$Xe. It is confirmed by a direct measurement with $^{37}$Ar isotope (2.82 keV) that the ionization yield is growing up with the energy decrease in the energy range below ~ 10 keV accordingly to the NEST predictions. Decay time of scintillation at 2.82 keV is measured to be $\tau = 25 \pm 3$ ns at the electric field of 3.75 kV/cm.

---

[*] Corresponding author.

# Contents

## 1. Introduction

The two-phase (liquid/gas) emission detector was invented about forty years ago [1],[2]. Through a long history of development it has become a powerful instrument for the experiments searching for rare processes. Recent rapid progress in the dark matter search experiments (ZEPLIN, XENON, and LUX [3],[4],[5]) was made with this type of detector. This was possible because of the unique performances of the two-phase detection technique. These performances include: superior sensitivity to low energy ionization, particle identification, 3D position reconstruction capability, and "wall-less" design benefiting from self-shielding.

Recently a two-phase emission detector has been proposed as a tool for discovery of a rare process of coherent neutrino-nucleus elastic scattering (CNNS) [6],[7],[8],[9]. This process is predicted by the Standard Model of electroweak interaction. However, it has not been observed yet, because a detector must be sensitive to the very low energy depositions from nuclear recoils.

Unfortunately, the detection properties of liquid noble gases which are used as targets for rare interactions are not well studied in the low-keV and sub-keV regions. The ionization yield for the low-energy nuclear recoils has been measured by the research groups who develop dark matter detectors. However, so far there were no direct experimental measurements of the ionization yield of liquid xenon (LXe) for electron recoils at the energies below 122 keV. An average energy to produce electron ion pair ($W_i$ value) that is equal to an inverse value of the ionization yield at the infinite electric field was measured in [10],[11] for 0.55 and 0.976-MeV $^{207}$Bi conversion electrons and in [12] for 122-keV gammas. The study of ionization yield of an argon filled two-phase detector in the low- and sub- keV region with $^{37}$Ar (0.27 and 2.82 keV) and $^{56}$Fe (5.9 keV) isotopes was recently reported in [13].

Here we report on first experimental study with a two-phase emission detector of ionization yield of electron recoils with energies below 100 keV produced in liquid xenon by the isotopes: $^{37}$Ar, $^{83m}$Kr, $^{241}$Am, $^{129}$Xe, $^{131}$Xe.

## 2. Detector

The emission detector used in this study was originally designed as a prototype of the ZEPLIN-III dark matter detector [14]. The active volume is a disk of 22 mm thickness and 105 mm diameter which contains 0.6 kg of liquid xenon with the total amount of LXe contained in the detector of ~ 5 kg. Seven 1"-diameter PMTs are immersed into LXe and view active volume from the bottom as shown in figure 1 [9]. An electrode structure includes a wired cathode, an intermediate field-shaping ring, and a flat aluminium-coated stainless steel anode. Part of our measurements (see below) was done with an $^{241}$Am source situated on the anode as shown in figure 1. There is a screening grid located in the vicinity of the PMT photocathodes to protect them from the applied electron drift field. The grid is biased at the same voltage as photocathodes. The gas gap where electroluminescence is generated is formed between the surface of liquid and the anode and has thickness of 5 mm. The nominal voltage between the anode and the cathode is 11.85 kV that corresponds to drift field 3.75 kV/cm in liquid and a maximal drift time of 9 μs. Maximal applied voltage is 15 kV. The voltage is supplied to the electrode system via Ceramaseal feedthroughs.

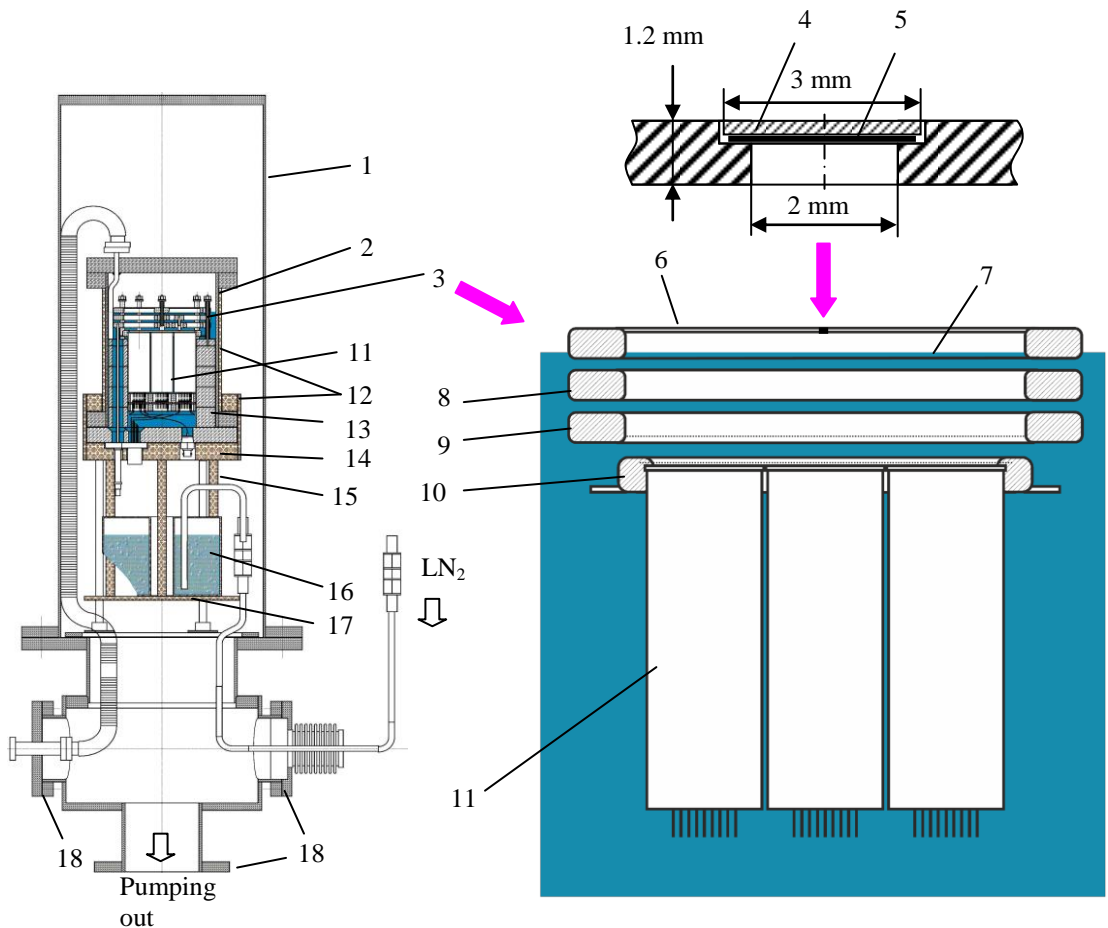

**Figure 1.** The emission detector: 1 – warm vessel; 2 – cold vessel; 3 – electrode system; 4 – $^{241}$Am source; 5 – stainless steel foil (200 μm); 6 – anode; 7 – LXe surface; 8 – field shaping ring; 9 – wired cathode (0.1 mm diam., 1 mm pitch); 10 – wired grid (0.1 mm diam., 1 mm pitch); 11 – PMTs; 12 – copper jacket; 13 – stainless steel LXe displacers; 14 – copper base plate; 15 – thermo conducting copper fingers; 16 – container filled with liquid nitrogen; 17 – copper base plate; 18 – vacuum ports.

In a two-phase emission detector [2], there are two signals produced by a particle interacting within the active volume of the noble liquid. The 1-st signal is primary scintillation (S1). The 2-nd one is ionization measured by means of electroluminescence (S2) in the gas phase of the detector. The charge is extracted from the liquid to the gas phase by applied electric field. Electron drift velocity and the distance between the point of interaction and the LXe surface define the time interval between the two signals.

The 175 nm wavelength Xe scintillation and electroluminescence photons are detected with FEU-181 photomultipliers equipped with $MgF_2$ windows and multi-alkali photo cathodes having 15% quantum efficiency at 200 nm wavelength. According to our previous measurements, no temperature dependence of the quantum efficiency in the range between room temperature and -120°C has been observed within 5% experimental accuracy. All seven PMTs are biased with the common voltage divider installed outside the detector.

Signals from PMTs were amplified by fast preamplifiers [15] installed on the cryostat. Then signals were split into high sensitivity and low sensitivity channels in order to obtain wider dynamic range. Signal in high sensitivity channels were post amplified with 8-channel Phillips Scientific 772. In addition, the preamplifiers are equipped with a remote control switches to select gain 0.5 or 5. All this allows one to measure the signal in the range from parts of single photoelectron (SPE) to ~ $5 \cdot 10^5$ SPE. Signals are digitised at 2 ns sampling rate (Struck SIS3350 12-bit) in the high sensitivity channels and 4 ns rate (CAEN V1720 12-bit) in the low sensitivity ones. Similar (but with some differences) structure of electronics was used for the old dataset (see below). Home developed amplifiers [16] with a fixed gain of 5 were used for pre- and post- amplification. LeCroy LT344 4-chanel 500 MHz and Tektronix 4- channel 625 MHz 8-bit oscilloscopes were used for digitizing in the high sensitivity channel and Struck SIS3320 (250 MHz, 12-bit) in the low sensitivity channel. Waveforms were recorded in time intervals of 10 μs before and after trigger point (20 μs in total).

## 3. Calibration lines

The experimental data were obtained using two datasets of measurements with calibration lines presented in table 1. The old dataset, in which we used $^{241}$Am and $^{83m}$Kr sources, is named RUN2009, the new one, RUN2013. To combine the results of both datasets together we used $^{83m}$Kr line as a reference point in both series of measurements.

Table 1. Energy lines and corresponding radioactive sources.

| Energy, keV | 2.82 | 13.95 | 17.75 | 30 | 40 | 41.5 | 59.5 | 80 | 662 |
|---|---|---|---|---|---|---|---|---|---|
| Isotope | $^{37}$Ar | $^{241}$Am | $^{241}$Am | $^{241}$Am | $^{129}$Xe | $^{83m}$Kr | $^{241}$Am | $^{131}$Xe | $^{137}$Cs |
| Description | EC, Auger | γ | γ | escape peak | n-γ | IC, γ, Auger | γ | n-γ | γ |
| RUN2009 | − | + | + | + | − | + | + | − | − |
| RUN2013 | + | − | − | − | + | + | − | + | + |

## 3.1 $^{37}$Ar

The isotope $^{37}$Ar decays to $^{37}$Cl with a half-life of 35 days by virtue of K- and L-capture with subsequent emission of Auger electron cascades with total energies 2.82 and 0.27 keV and intensities ~10:1, correspondingly [17]. The $^{37}$Ar was produced by irradiation of the natural Ca (97% of $^{40}$Ca) by 14-MeV neutrons from the d-t generator in the reaction:

$$^{40}\text{Ca} + n \rightarrow {}^{37}\text{Ar} + {}^{4}\text{He}$$

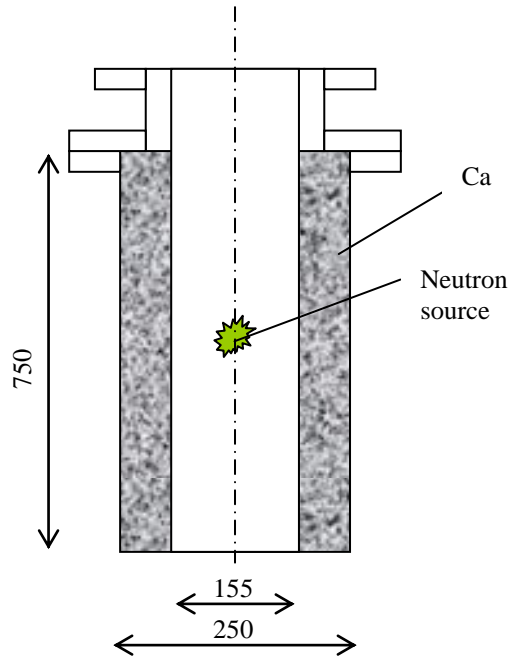

**Figure 2.** Container with natural Ca to expose by fast neutrons. Dimensions are in mm.

The natural Ca in a form of chips from a turning machine was encapsulated in a hermetic container as shown in figure 2. The chips were prepared by machining of the calcium cylinder in an argon atmosphere to prevent oxidation and hydration of metallic calcium. The container was pumped out before irradiation during several days with heating to ~ 200°C and controlling the residual outgassing with an RGA. The main residual gas was hydrogen originated from the hydration reaction which took place when the calcium chips were exposed to an open air during filling the container. After outgassing, the container was isolated from air by vacuum valves and was disconnected from the gas system. Then it was shipped to a stationary d-t neutron generator, where it was exposed to $5 \cdot 10^{12}$ neutrons in $4\pi$. Using MCNP simulation we estimated that about $10^{10}$ atoms of $^{37}$Ar had been produced. Then the container was connected to the gas system again and baked for one week at ~ 200°C to extract the argon atoms from the calcium. After that, the container was filled with Xe gas. The RUN2013 experimental run was divided into two parts: RUN2013-1 and RUN2013-2. In the 1$^{st}$ one, we used the xenon free from the $^{37}$Ar isotope; in the 2$^{nd}$ one, mixed with $^{37}$Ar. To prepare this mixture we condensed all the gas from the detector after the 1$^{st}$ part of the run to the cylinder cooled down with LN$_2$ and then condensed there the gas from the container. The detector was refilled again with the Xe + $^{37}$Ar mixture. This procedure was made to ensure the total use of the produced $^{37}$Ar atoms.

## 3.2 $^{83m}$Kr

The $^{83m}$Kr radioactive gaseous isotope [18] decays with a half-life of $T_{1/2}$ = 1.83 h via internal conversion (IC) with energy release of 32.1 keV to the first $^{83}$Kr excited state (9.4 keV), which then decays via IC (95%) or emission of γ (5%) to the ground state with $T_{1/2}$ = 154 ns. The $^{83m}$Kr is constantly produced in a decay of $^{83}$Rb ($T_{1/2}$ = 86.2 d). The $^{83}$Rb isotope was produced at the Institute of Nuclear Researches RAS for the "Troitsk nu-mass" experiment [19]. The isotope was deposited on a stainless-steel foil and encapsulated in a stainless-steel container. This container was made from the standard MDC CF 2.75" nipple and was incorporated in the gas system ("$^{83m}$Kr–generator").

The $^{83m}$Kr was delivered into the detector active volume by a flow of Xe gas. For this purpose the part of the Xe was taken from the input of the detector chamber to a bellow hose bended in U-shape and cooled down by liquid nitrogen. After warming up the hose the Xe gas was coming to the "$^{83m}$Kr–generator" container heated to ~100°C for several hours beforehand in order to release the $^{83m}$Kr from Rb. Then the Xe gas was condensed back to the detector chamber through Millipore Waferpure Micro gas purifier. In order to ensure the same level of LXe in the detector chamber as before injection of $^{83m}$Kr into detector the process of condensation is stopped when the residual pressure in the system (including the "$^{83m}$Kr–generator") is equal to the initial one.

In the detector, the decay of $^{83m}$Kr produces two fast sequential scintillation signals ($S1_1$ and $S1_2$) corresponding to the energies 32.1 and 9.4 keV and one wide electroluminescent signal (S2) corresponding to the total deposited energy 41.5 keV.

In the RUN2013, the $^{83m}$Kr was injected twice: before and after $^{37}$Ar injection.

## 3.3 $^{137}$Cs and $^{241}$Am

We used $^{137}$Cs (662 keV, in RUN2013) gamma-source to do measurements at high energies. The source was installed outside of the detector cryostat at the level of active volume.

The $^{241}$Am source was installed at the mirror anode of the electrode structure (in RUN2009). The source was deposited on a stainless steel disc with a diameter of 3 mm (4 in figure 1) and was protected with $TiO_2$ 1-μm layer. In order to screen the alpha particles there was a stainless steel foil in front of the source (5 in figure 1). Americium-241 undergoes alpha decay to $^{237}$Np. The major gamma-line of 59.54 keV is produced from de-excitation of the exited level of $^{237}$Np. The decay is accompanied with Neptunium *L* X-rays, the most intense of which are: $L_{α1}$ and $L_{β1}$, 13.95 and 17.75 keV correspondently [20]. The 59.54-keV gamma interacts with the *K*-shell electron of Xe atom. The vacancy on *K*-shell is filled by transition of electron from the *L*-shell with emission of a characteristic photon with 29.5-keV energy. Since interaction of 59.54-keV gamma with the LXe takes place within 1 mm from the surface, there is the large probability for this X-ray to escape the LXe resulting in the total energy deposition in the LXe of 30.0 keV.

## 3.4 $^{129}$Xe, $^{131}$Xe

For calibration of the LXe we also used 40 and 80 keV lines obtained from n-gamma reaction on $^{129}$Xe and $^{131}$Xe, correspondingly. Fast neutrons for this study were produced by Pu-Be source installed outside the cryostat.

## 4. Data analysis

### 4.1 Signal reconstruction

Recorded signal waveforms for each channel are examined for pulses. All pulses: S1, S2, single photoelectrons (SPE), noise or any pulses of unknown origin that exceed threshold are counted and parameterized. The latter includes calculation of amplitude, area, time and width of pulse. Threshold is calculated as a multiple of baseline noise and is low enough to catch even single photoelectrons. Pulses in different channels that coincide within the defined time interval are considered to form one cluster. The S1 and S2 clusters are identified by width and shape parameters of the pulses. Area value for clusters was calculated as sum of areas of individual pulses within cluster. The S2 area values were corrected then for finite electron lifetime in liquid. Both S1 and S2 area values were corrected also with a spatial XY correction algorithm [21] (called below as 'geometry correction'). This correction takes into account the radial dependence of light collection. The light response functions required for geometry correction was taken from the $^{83m}$Kr data analysis. Event energy was calculated using the area of S2 cluster.

We also analyzed 'single electron' events (SE) [8],[22],[23],[24],[25]. Such events are extremely small and look like a sequence of SPE signals with a total duration of ~ 1 μs that is typical to the width of S2 signals. We used a dedicated trigger to obtain a clear sample of such events. A trigger was made to select the events with number of SPEs greater than 2 and appeared without correlation with any large signal. Such events are also called spontaneous SE (see [8],[25]). We select these evens as sequence of SPEs taking place in a time window equal to the width of electroluminescence signal. Additional cuts were applied to remove events that actually are low-scintillation signals.

### 4.2 Free electron lifetime

Electron lifetime of free electrons in LXe is a very important parameter of a two-phase emission detector. In the experimental runs, this value was quite moderate, about 15 μs on average at the 3.75 kV/cm operation electric field. That is ~1.5 times greater than the full drift time between the cathode and the LXe surface. Thus, correction of the S2 values is required in order to obtain the true value of ionization signal. Since the lifetime was changing slowly during RUN2013, we used two methods to monitor it. First, we analyzed of the shape of cosmic muon signals. Muons produce constant ionization at all depth from surface till very bottom. So, resulting electroluminescence signal should decay along drift time with a characteristic lifetime. Second, we calculated the lifetime directly from fit of distribution of the S2 peak vs $T_{dr}$ data, where $T_{dr}$ is the interval between the S1 and S2 signal, a drift time of electrons between the interaction point and the LXe surface.

### 4.3 Analysis of the $^{83m}$Kr data

The $^{83m}$Kr events were selected from the RUN2013-1 dataset using a criterion of presence of two scintillation pulses ($S1_1$ and $S1_2$) following one by another and single S2 pulse with additional condition of $S1_1 > S1_2$.

To check if the $S1_1$ and $S1_2$ pulses really correspond to the $^{83m}$Kr decay we plotted the distribution of the intervals between them. The $T_{1/2}$ decay time of the exponent is 132 ± 21 ns that within the error bars corresponds to a table value of 154 ns for the half-life of excited level. This confirms the correct selection of the $^{83m}$Kr. The S2 signals were corrected then for lifetime and geometry. The resulted S2 distribution is shown in figure 3 in red color.

## 4.4 Analysis of the $^{37}$Ar data

After filling the detector with Xe+$^{37}$Ar mixture (RUN2013-2), the count rate of the gamma-trigger raised up from the original ~15 Hz to ~25 Hz. We attributed this to the appearance of the main 2.82-keV line of $^{37}$Ar. In the $^{37}$Ar data analysis, the peak position was defined as a relative position to the $^{83m}$Kr peak obtained in RUN2013-1. However, there was a significant difference in event selection conditions related to the $^{37}$Ar and $^{83m}$Kr data. For the $^{37}$Ar events, S1 is single and much smaller than in the $^{83m}$Kr ones. Its value is at a level of SPE. In order to check that in our event selection we take the real S1 of $^{37}$Ar rather than an accidental SPE, we calculated electron lifetime from the $^{37}$Ar data. We found this value to be equal to 15 ± 8 μs which is consistent with that obtained by other methods. Another criterion was obtaining the right decay time of scintillation. To obtain the decay time we selected the events having two SPE pulses (not necessarily from the same PMT) before the S2 signal and plotted the distribution of the intervals between them (see figure 4). One or both of them may belong to a real scintillation or may be accidental. In the case when both SPE signals are those from the real scintillation the decay times in the distribution must be close to the lifetimes of exited states of Xe$_2$* molecules. The value of τ = 25 ± 3 ns (single exponent) obtained from the distribution shown in figure 4 is consistent within the error bars with τ = 28.1 ± 0.6 ns obtained in [26] for gammas from 40 to 511 keV and with τ = 27 ± 1 ns obtained in [27] for MeV-gammas from $^{207}$Bi measured at the same electric field. This decay time is defined by the lifetime of the $^3\Sigma_u^+$ state of Xe$_2$*. The S2 signals were corrected with the use of the electron lifetime, and then geometry correction was applied to them. The resulted S2 area distribution is shown in figure 3 in blue color. The obtained ratio of $^{83m}$Kr to $^{37}$Ar peak position is 13.75 ± 1.90.

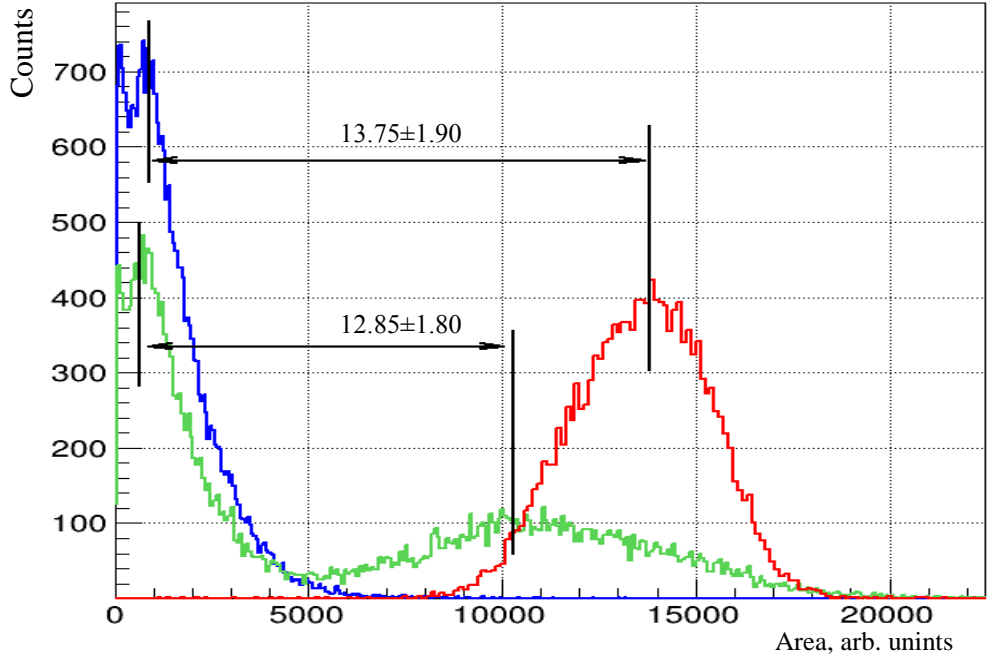

**Figure 3.** Distribution of S2 signal areas of $^{37}$Ar $^{83m}$Kr events; *red* - events from the $^{83m}$Kr runs, *blue* – events from $^{37}$Ar runs, *green* – events from the runs with both $^{37}$Ar and $^{83m}$Kr in the detector (without lifetime correction).

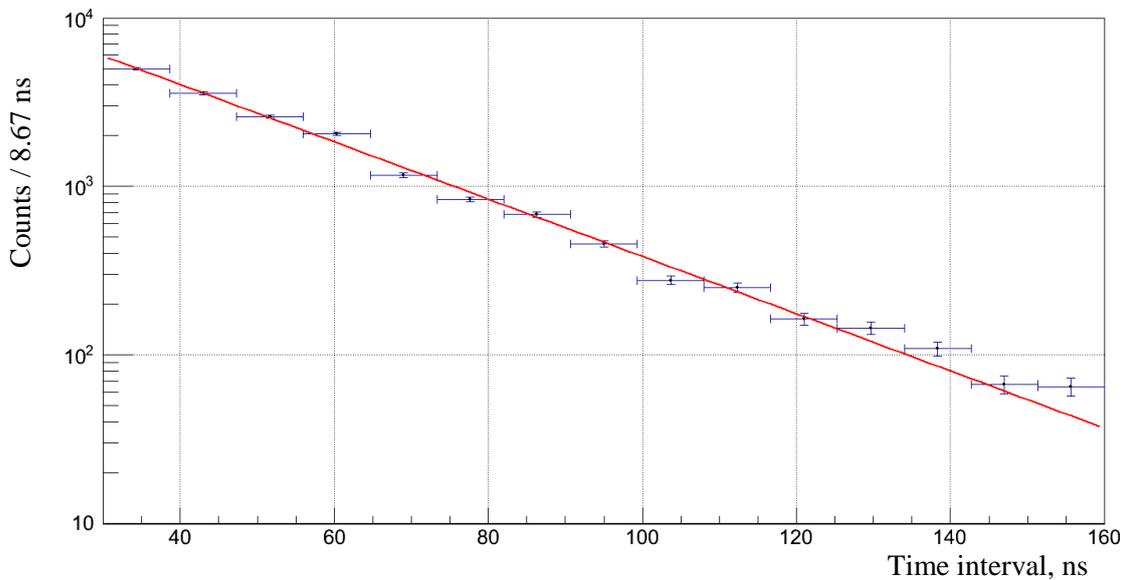

**Figure 4.** Distribution of time intervals between the 1$^{st}$ and the 2$^{nd}$ SPE pulses, attributed to the scintillation of $^{37}$Ar. Exponential fit corresponds to a decay time of scintillation τ = 25 ± 3 нс.

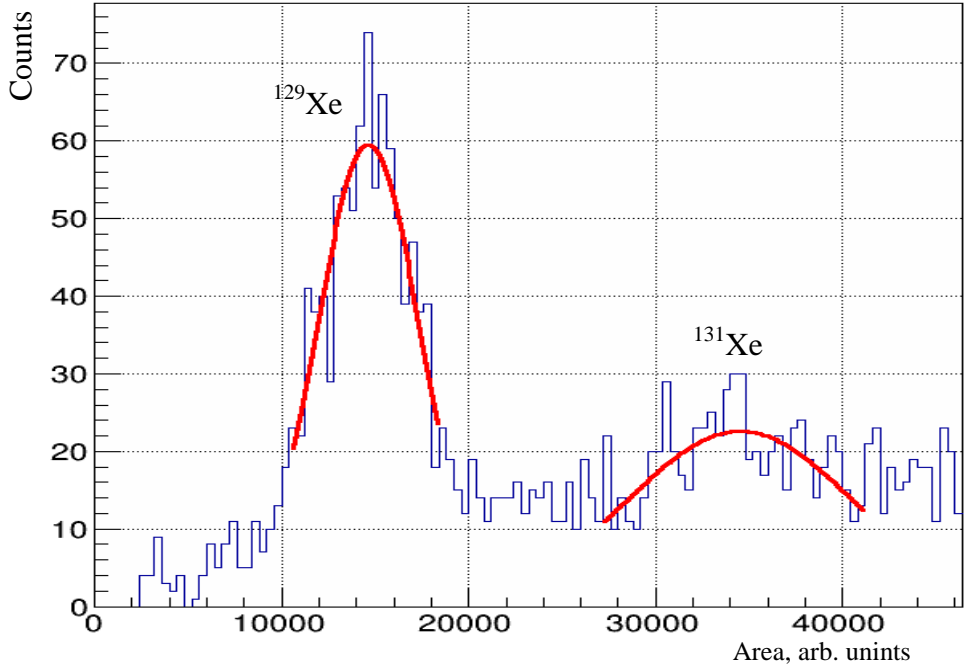

**Figure 5.** Distribution of S2 signal areas of the events from PuBe source.

In addition, in order to reduce systematics we analyzed the datasets where both the $^{37}$Ar and the $^{83m}$Kr events were collected simultaneously. We did not apply lifetime correction to these data but used only the geometry correction (see green histogram in figure 3). In this case the peaks are broader, and their positions are lower in compare with the corrected peaks. The obtained ratio of $^{83m}$Kr to $^{37}$Ar peak position is 12.85 ± 1.80.

We used for the final ratio the weighted mean value of 13.3 ± 1.3.

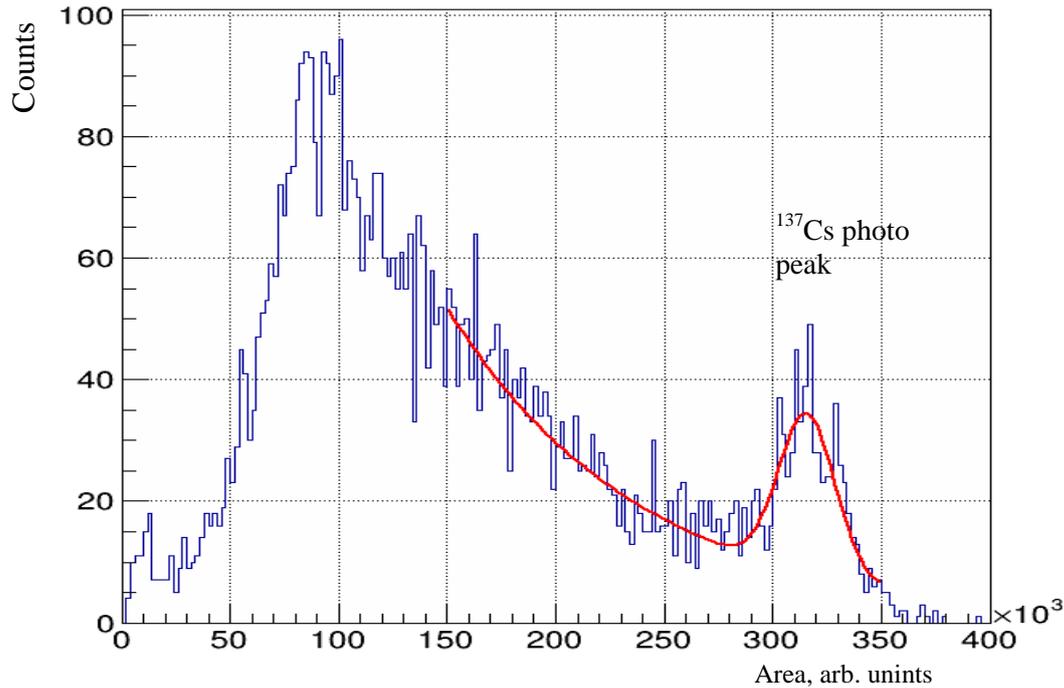

**Figure 6.** Distribution of S2 signal areas of the events from $^{137}$Cs source. The right part of the histogram is fitted by Gaussian on exponential substrate.

### 4.5 Analysis of the other points of RUN2013

The Pu-Be and $^{137}$Cs experimental data were acquired with the preamplifiers gain of 0.5 in order to fit to the dynamic range of electronics to the high values of the S2 signals correspondent to 80 and 662 keV energies. The data were corrected using the lifetime and geometry correction algorithms. The resulted S2 distributions are given in figure 5 and figure 6 for Pu-Be and $^{137}$Cs, correspondingly.

### 4.6 Analysis of RUN2009 data

The energy spectrum obtained with $^{241}$Am is shown in figure 7. The energy spectrum obtained with $^{83m}$Kr source is also given in this figure. Since the $^{241}$Am source was installed in the middle of the anode it predominately illuminated the central part of the detector. In order to keep the similar conditions for the $^{83m}$Kr events we selected only those of them that took place within the radius of 20 mm. No geometrical corrections were applied for both datasets. The $^{83m}$Kr were corrected for the lifetime of 17.9 ±1.8 μs. No lifetime corrections were applied to the $^{241}$Am events because all of them had a shallow depth of an order of 1 mm. The lines 13.95 keV (Np $L\alpha_1$) and 17.75 keV (Np $L_{\beta 1}$) are not resolved each from other. The intensity of the 1st line is strongly reduced by absorption in the stainless steel foil. This was taken into account at fitting procedure.

### 4.7 Error estimation

Several sources of uncertainties of the obtained experimental points were analyzed. The uncertainty caused by the event selection procedure was estimated by variation of the selection criteria such as S1 and S2 duration and interval between them. The uncertainty of the S2 values coming from the selection of fit intervals was estimated by variation of these intervals. One of the main contributions to the uncertainty comes from the reconstruction procedure, and is caused by the uncertainty of the electron lifetime. It was estimated by variation of the lifetime value. Moreover, for the reconstruction of the detector response to the 662-keV gammas, the data obtained in the low sensitivity channel (see above) was used. Uncertainties of scale factors between the high and low sensitivity individual channels also affect on the final accuracy.

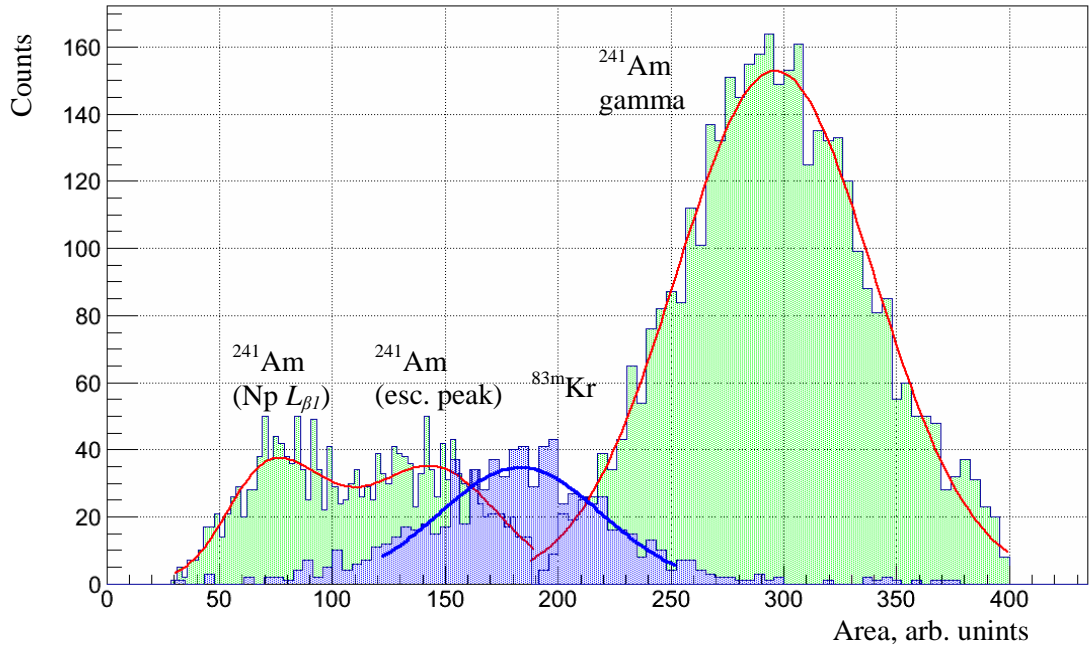

**Figure 7.** Distribution of S2 signal areas of the events from $^{241}$Am and $^{83m}$Kr sources in the RUN2009 experimental run.

The uncertainty of the $^{37}$Ar point is combined from the uncertainties of the $^{83m}$Kr point and the ratio between $^{37}$Ar and $^{83m}$K signals.

### 4.8 Ionization yield

The experimental data on specific ionization yield in e$^-$/keV for electron recoils in the liquid xenon obtained from the peak positions are given in figure 8 and in table 2.

Table 2. Specific ionization yield for electron recoils in the liquid xenon at different energies.

| Source | Energy, keV | Ionization yield, e$^-$/keV |
| --- | --- | --- |
| $^{37}$Ar | 2.82 | 47.8 ± 5.5 (syst.) |
| $^{241}$Am, Np $L_{\beta 1}$ | 17.75 | 44.3 ± 2.6 (stat.) $^{+1.3\,(syst.)}_{-4.0\,(syst.)}$ ± 2.9* |
| $^{241}$Am, esc. peak (59.5-29.5 keV) | 30 | 48.0 ± 0.7 (stat.) $^{+0.7\,(syst.)}_{-1.2\,(syst.)}$ ± 3.1* |
| $^{129}$Xe | 40 | 47.3 ± 0.4(stat.) ± 1.4 (syst.) |
| $^{83m}$Kr | 41.5 | 43.2 ± 0.1 (stat.) ± 2.6 (syst.) |
| $^{241}$Am | 59.5 | 48.9 ± 0.1(stat.) ± 0.3 (syst.) ± 3.2* |
| $^{131}$Xe | 80 | 52.1 ± 1.4 (stat.) $^{+1.1\,(syst.)}_{-1.6\,(syst.)}$ |
| $^{137}$Cs | 662 | 62.26 ± 0.2 (stat.) $^{+3.0\,(syst.)}_{-3.6\,(syst.)}$ |

* comes from the uncertainty of the RUN2009 to RUN2013 scale factor, marked by ⌶ bars in figure 8; the uncertainty that comes from the W$_i$ error bars is not shown

The data on ionization yield at the 3.75 kV/cm electric field are obtained in a relative scale without absolute normalization since we didn't do any absolute calibration of the ionization signals. In order to normalize the data we used an empirical model of ionization yield for electrons and gammas given in [28]. In this model, the charge extracted from electron track (the portion of electrons that escaped recombination) is described by the formula:

$$Q(E) = \frac{Q_\infty}{(1+k/E)}.$$

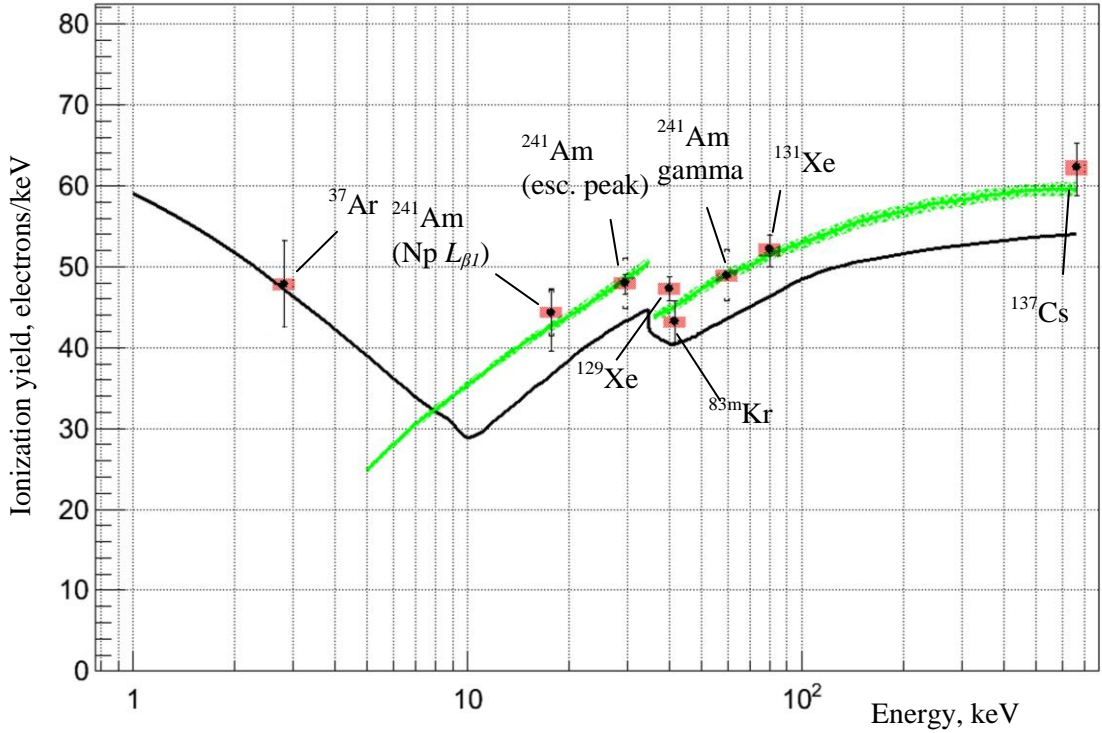

**Figure 8.** Experimental data for the ionization yield of LXe for electron recoils. See text for explanation.

This formula is generally accepted for parameterization of ionization saturation curves, i.e. dependences of a fraction of the charge that escape recombination to the total one (see [11], [29]). In this formula, $E$ is electric field strength, $Q_\infty = 1/W_i$, $W_i$ is ionization work function, $k$ is value characterizing recombination process. In the reference [28], the only dependences of $k$ values versus the gamma-ray or electron energies were derived from the measured ionization saturation curves for the low energy gammas (15.3, 17.3, 21.4 keV) without absolute normalization of the ionization yield. With the use of the dependence for gammas we obtained the green line in figure 8. This curve is normalized in vertical direction using $W_i = 15.1 \pm 0.2$ eV which is the weighted average between the values obtained in [10],[12], and [30] (see the next section). Vertical uncertainty of this curve (shown by shadowed green) comes from the $W_i$ error bars. Our data points (except of $^{37}$Ar one) are fitted to this curve with the use of vertical scale factor as a free parameter. The same relative uncertainty goes to the experimental points (shown by red shadowing, the width of this shadowing in horizontal direction is arbitrary). On the same plot, the curve produced by NEST (Noble Element Simulation Technique) [31] for the 3.75 kV/cm electric field is shown.

Normalization of the data points in vertical using $W_i = 15.1 \pm 0.2$ eV results in a photoelectron yield per one single ionization electron (SPE/SE) of $9.83 \pm 0.35$. To obtain this value we used the weighted averaged $0.78 \pm 0.025$ of the liquid-to-gas electron extraction coefficient from [32],[33],[22]. Since in [33] there is no error bars specified for the experimental points we've taken for them the same errors as in [32]. In order to obtain the SPE/SE value by different way we've carried out a special dedicated study of single electron events in RUN2013. The most probable SPE/SE number obtained by counting SPEs within the SE sequence is equal to 7.3 ± 1.1. Note, that these spontaneous SE events are distributed nearly uniformly in the horizontal plane [8],[34]. Therefore, the obtained SPE/SE value is underestimated since a light response function decreases to periphery. After correction this value with the use of the light response function obtained from $^{83m}$Kr calibration we obtained SPE/SE = 8.5 ± 1.2. This value agrees within the error bars with that one obtained by normalization.

## 5. Discussion

One can see the remarkable coincidence of both curve shapes in figure 8 in the region of energies higher than 10 keV. The experimental data fit very well to the general behavior of both of them. However, the NEST curve goes lower than the curve plotted with the model given in [28], normalized with the use of $W_i = 15.1 \pm 0.2$ eV.

This absolute normalization of experimental points is rather difficult and might be debated. There are several measurements of $W_i$-value in literature. Historically, the first reliable measurement of $W_i = 15.6 \pm 0.3$ eV was done in [10]. In that work, the $W_i$-value in the liquid Xe was measured for MeV electrons and gammas from $^{207}$Bi with respect to the $W_i$-value in gas that was well established before. Quite close to this but higher $W_i$-value equal to 16.5 ± 0.8 eV was obtained for the 122-keV gammas in the dark matter detector ZEPLIN III [12] with the similar two-phase emission technique as is used in the current study. Absolute calibration was based on knowledge of the SPE/SE value and measured saturation curve. In the work [30], a universal value $W = 13.46 \pm 0.29$ eV to create a quantum that is an ionization electron or a scintillation photon was measured with the use of a double-phase emission technique. Absolute calibration of both scintillation and ionization scales was done with the use of a charge sensitive preamplifier (see [30] for details). The corresponded $W_i$-value can be obtained from $W$ with the use of the exiton-to-ion ratio of 0.06 for LXe (see [10],[35],[36]): $W_i = (1+0.06) \cdot W = 14.27 \pm 0.3$ eV. We used these three mentioned values to obtain the weighted average. There are, however, the results of other measurements that we do not take for averaging. In [37], $W = 14$ eV was obtained with a two-phase detection technique. Taking into account the exiton-to-ion ratio that corresponds to $W_i = 14.84$. This value is not used because no error bars are given to this value. In [38], $W_i = 13.6 \pm 0.2$ eV was obtained. We did not use it because the universal $W$-value derived from it as $W = W_i/1.06 = 12.8 \pm 0.2$ eV is extraordinary low with respect to the set of the other experimental $W$-values (see compilation of data in [31]). Agreement within the error bars of the SPE/SE value obtained experimentally with that derived from normalization using $W_i = 15.1 \pm 0.2$ eV confirms the correctness of normalization.

However, the offset between the NEST curve and experimental points at the high energies is less important for us in compare with the fact that the low-energy $^{37}$Ar point does follow to the prediction of the NEST but not to the model given in [28]. The last one is based on the common knowledge that the recombination goes stronger at higher ionization density. However, in the low energy range (< ~ 10 keV), the recombination probability depends on the energy via the number of ions but not on dE/dx (Thomas-Imel regime [39]). For these energies, the particle tracks are smaller than the thermalization distance of ionization electrons, thus only part of the electrons may recombine with the ions on track. This approach was successfully applied by Dahl to explain the odd drop in the scintillation light yield at low energies [36]. The NEST code is based on two different models depending on energy: Thomas-Imel and Doke-Birks (see details in [31]).

## 6. Conclusion

The experimental data on specific ionization yield for electron recoils in liquid xenon were obtained for the first time in the energy region below 100 keV. For the energies higher than ~ 10 keV the data demonstrates excellent agreement with the classical models that implies decreasing of the yield with the energy decrease and with Thomas-Imel model at lower energies. Our $^{37}$Ar point is the first direct experimental evidence that in liquid xenon the ionization yield for electron recoils increases with energy decrease similarly as this was demonstrated recently for liquid argon [13]. This demonstrates that a noble gas two-phase emission detector is sensitive to the very low-energy particles.

In addition to ionization measurements, we've obtained the decay time of scintillation in liquid xenon at 2.82 keV ($^{37}$Ar): $\tau = 25 \pm 3$ ns. This corresponds to the decay of $^{3}\Sigma_{u}^{+}$ state of Xe$_2$*.

## Acknowledgments

This study was supported RF Government under contracts of NRNU MEPhI with the Ministry of Education and Science of №11.G34.31.0049 from October 19, 2011 № P881 from May 26, 2010 and by the Russian Foundation for Basic Research under the contract of №11-02-00668-a.

We also grateful to the Institute for Nuclear Researches RAS and to B.L. Zhuikov personally for production of the $^{83}$Rb.

We are grateful to M. Szydagis and H. Araújo for the very valuable discussions on NEST code. We are thankful to Ben Rybolt for careful reading of the manuscript.

## References

[1] B.A. Dolgoshein, V.N. Lebedenko, and B.U. Rodionov, New method of registration of tracks of ionizing particles in condensed matter, *JETP Lett.* **11** iss. 11 (1970) 351.

[2] A. Bolozdynya, "Emission detectors," Singapore: World Scientific, 2010, pp. 107-117.

[3] D.Yu. Akimov, H.M. Araújo, E.J. Barnes et al., WIMP-nucleon cross-section results from the second science run of ZEPLIN-III, *Phys. Lett.* B **709** (2012) 14.

[4] E. Aprile, M. Alfonsi, K. Arisaka et al., Limits on spin-dependent WIMP-nucleon cross sections from 225 live days of XENON100 data, *Phys. Rev. Lett.* **111** iss. 2 (2013) 021301.

[5] D.S. Akerib, H.M. Araujo, X. Bai et al., First results from the LUX dark matter experiment at the Sanford Underground Research Facility", *Phys. Rev. Lett.* **112** (2014) 091303 [arXiv:1310.8214].

[6] C. Hagmann, A. Bernstein, Two-phase emission detector for measuring coherent neutrino-nucleus scattering, *IEEE Trans. Nucl.* Sci. **51** (2004) 2151 [nucl-ex/0411004].

[7] D. Akimov, A. Bondar, A. Burenkov, and A. Buzulutskov, Detection of reactor antineutrino coherent scattering off nuclei with a two-phase noble gas detector, *JINST* **4** (2009) P06010 [arXiv:0903.4821].

[8] E. Santos, B. Edwards, V. Chepel et al., Single electron emission in two-phase xenon with application to the detection of coherent neutrino-nucleus scattering, *JHEP* **1112** (2011) 115 [arXiv:1110.3056].

[9] D.Yu.Akimov, I.S. Alexandrov, V.I. Aleshin et al., Prospects for observation of neutrino-nuclear neutral current coherent scattering with two-phase Xenon emission detector, *JINST* **8** (2013) P10023 [arXiv:1212.1938].

[10] T. Takahashi, S. Konno, T. Hamada et al., Average energy expended per ion pair in liquid xenon, *Phys. Rev.* **A12** (1975) 1771.

[11] E. Shibamura, A. Hitachi, T. Doke et al., Drift velocities of electrons, saturation characteristics of ionization and W-values for conversion electrons in liquid argon, liquid argon-gas mixtures and liquid xenon, *Nucl. Instr. Meth.* **131** (I975) 249.

[12] M. Horn, V. A. Belov, D. Yu. Akimov et al., Nuclear recoil scintillation and ionisation yields in liquid xenon from ZEPLIN-III data, *Phys. Lett.* **B705** (2011) 471 [arXiv:1106.0694].

[13] S. Sangiorgio, A. Bernstein, J. Coleman et al., First demonstration of a sub-keV electron recoil energy threshold in a liquid argon ionization chamber, *Nucl. Instrum. Meth.* **A728** (2013) 69 [arXiv:1301.4290].

[14] D.Yu. Akimov, G.J. Alner, H.M. Araujo et al., The ZEPLIN-III dark matter detector: instrument design, manufacture and commissioning, *Astropart. Phys.* **27** (2007) 46 [astro-ph/0605500].

[15] D.S. Akerib, X. Bai, S. Bedikian et al., The Large Underground Xenon (LUX) Experiment, *Nucl. Instrum. Meth.* **A704** (2013) 111 [arXiv:1211.3788].

[16] D. Yu. Akimov, Yu. K. Akimov, A. A. Bogdzel' et al., A Low-Noise Fast Eight-Channel Preamplifier, *Instr. Experim. Tech.* **45** iss. 2 (2002) 207.

[17] V. I. Barsanov, A. A. Dzhanelidze, S. B. Zlokazov et al., Artificial neutrino source based on the Ar-37 isotope, *Phys. Atom. Nucl.* **70** (2007) 300.

[18] L. W. Kastens, S. B. Cahn, A. Manzur, and D. N. McKinsey, Calibration of a liquid xenon detector with $^{83}Kr^m$, *Phys. Rev.* **C80** (2009) 045809 [arXiv:0905.1766].

[19] A. I. Belesev, E.V. Geraskin, B.L. Zhuikov et al., Investigation of space-charge effects in gaseous tritium as a source of distortions of the beta spectrum observed in the Troitsk neutrino-mass experiment, *Phys. Atom. Nucl.* **71** (2008) 427.

[20] M.C. Lepy, J. Plagnard, L. Ferreux, Measurement of (241)Am L X-ray emission probabilities, *Appl. Radiat. Isot.* **66** (2008) 715.

[21] V.N. Solovov, V.A. Belov, D.Y. Akimov et al., Position Reconstruction in a Dual Phase Xenon Scintillation Detector, *IEEE Trans. Nucl. Sci.* **59** (2012) 3286 [arXiv:1112.1481].

[22] E. Aprile, M. Alfonsi, K. Arisaka et al., Observation and applications of single-electron charge signals in the XENON100 experiment, *J. Phys. G: Nucl. Part. Phys.* **41** (2014) 035201 [arXiv:1311.1088].

[23] B. Edwards, H.M. Araújo, V. Chepel et al., Measurement of single electron emission in two-phase xenon, *Astropart. Phys.* **30** (2008) 54 [arXiv:0708.0768].

[24] A.A. Burenkov, D.Yu. Akimov, Yu.L. Grishkin et al., Detection of a single electron in xenon-based electroluminescent detectors, *Phys. Atom. Nucl.* **72** (2009) 653.

[25] D.Yu. Akimov, I.S. Aleksandrov, V.A. Belov et al., Measurement of single-electron noise in a liquid-xenon emission detector, *Instrum. Exp. Tech.* **55** (2012) 423.

[26] D. Yu Akimov, V. F. Batyaev, S. P. Borovlev et al., Liquid Xenon for WIMP searches: measurement with a two-phase prototype, *Proc. 4th Int. Workshop on the Identification of Dark Matter (IDM 2002)*, 2-6 Sep 2002. York, United Kingdom; ed. N. J. C. Spooner V. Kudryavtsev, Singapore: World Scientific, 2003, p. 371.

[27] S. Kubota, M. Hishida and J. Raun, Evidence for a triplet state of the self-trapped exciton states in liquid argon, krypton and xenon, *J. Phys.* **C11** (1978) 2645.

[28] T.Ya. Voronova, M.A. Kirsanov, A.A. Kruglov et al., Ionization yield from electron tracks in liquid xenon, *Sov. Phys. Tech. Phys.* **34** iss. 7 (1989) 825; transl. *Zh. Tekh. Fiz.* (in Russian) **59** (1989) 186.

[29] V. Chepel and H. Araújo, Liquid noble gas detectors for low energy particle physics, *JINST* **8** (2013) R04001 [arXiv:1207.2292].

[30] T. Shutt, C.E. Dahl, J. Kwong, A. Bolozdynya and P. Brusov, Performance and fundamental processes at low energy in a two-phase liquid xenon dark matter detector, *Nucl. Instrum. Meth.* **A579** (2007) 451 [astro-ph/0608137].

[31] M. Szydagis, N. Barry, K. Kazkaz et al., NEST: a comprehensive model for scintillation yield in liquid xenon, *JINST* **6** (2011) P10002 [arXiv:1106.1613] http://nest.physics.ucdavis.edu/site/.

[32] E. M. Gushchin, A. A. Kruglov, V. V. Litskevich et al., Electron emission from condensed noble gases, *Sov. Phys. JETP* **49** iss. 5 (1979) 856.

[33] E. Aprile, K.L. Giboni, P. Majewski et al., Proportional light in a dual-phase xenon chamber, *IEEE Trans. Nucl. Sci.* **51** iss. 5 (2004) 1986.

[34] D.Yu. Akimov, I.S. Alexandrov, V.A. Belov et al., A Two-phase Emission Liquid Xe Detector for Study of Low-Ionization Events at the Research Reactor IRT MEPhI, in press *IEEE Trans. Nucl. Sci.*

[35] T. Doke, Fundamental properties of liquid argon, krypton and xenon as radiation media, *Portugal Phys.* **12** (1981), 9.

[36] C.E. Dahl, The physics of background discrimination in liquid xenon, and first results from XENON10 in the hunt for WIMP dark matter, Ph.D. thesis, Princeton University, Princeton U.S.A. 2009.

[37] E. Aprile J. Angle, F. Arneodo et al., Design and performance of the XENON10 dark matter experiment, *Astropart. Phys.* **34** (2011) 679 [arXiv:1001.2834].

[38] I.M. Obodovskii and S.G. Pokachalov, Average ion pair formation energy in liquid and solid xenon, *Sov. J. Low Temp. Phys.* **5** (1979) 393.

[39] J. Thomas and D.A. Imel, Recombination of electron-ion pairs in liquid argon and liquid xenon, *Phys. Rev.* **A36** (1987) 614.